\documentclass{emulateapj}
\usepackage{natbib}
\usepackage{epsfig}
\usepackage{graphicx}
\usepackage{subfigure}
\usepackage{float}
\usepackage{amsmath}
\usepackage{color}
\usepackage{amssymb}
\usepackage{amsfonts}
\usepackage[colorlinks,linkcolor=blue,anchorcolor=green,citecolor=blue]{hyperref}
\usepackage{tikz,xcolor,hyperref}
\usepackage{footmisc}

\definecolor{lime}{HTML}{A6CE39}
\DeclareRobustCommand{\orcidicon}{
	\begin{tikzpicture}
	\draw[lime, fill=lime] (0,0)
	circle [radius=0.16]
	node[white] {{\fontfamily{qag}\selectfont \tiny ID}};
	\draw[white, fill=white] (-0.0625,0.095)
	circle [radius=0.007];
	\end{tikzpicture}
	\hspace{-2mm}
}
\foreach \x in {A, ..., Z}{%
	\expandafter\xdef\csname orcid\x\endcsname{\noexpand\href{https://orcid.org/\csname orcidauthor\x\endcsname}{\noexpand\orcidicon}}
}

\begin{document}
	\title{Repeating Fast Radio Bursts from Pulsar-Asteroid Belt Collisions: Frequency Drifting and Polarization}
	\author{Zenan Liu\altaffilmark{1,2}, Wei-Yang Wang\altaffilmark{3,4}, Yuan-Pei Yang\altaffilmark{5}, and Zi-Gao Dai\altaffilmark{1,2}}
	\affil{$^1$School of Astronomy and Space Science, Nanjing University, Nanjing 210023, China; dzg@nju.edu.cn}
	\affil{$^2$Key Laboratory of Modern Astronomy and Astrophysics (Nanjing University), Ministry of Education, Nanjing, China}
	\affil{$^3$Key Laboratory for Computational Astrophysics, National Astronomical Observatories, Chinese Academy of Sciences, 20A Datun Road, Beijing 100101, China}
	\affil{$^4$University of Chinese Academy of Sciences, Beijing 100049, China}
	\affil{$^5$South-Western Institute for Astronomy Research, Yunnan University, Kunming 650500, People's Republic of China}
	
	\begin{abstract}		
		Fast radio bursts (FRBs) are a new kind of extragalactic radio transients. Some of them show repeating behaviors. Recent observations indicate that a few repeating FRBs (e.g., FRB 121102) present time--frequency downward drifting patterns and nearly 100$\%$ linear polarization. Following the model of \citet{dai 2016} who proposed that repeating FRBs may originate from a slowly-rotating, old-aged pulsar colliding with an asteroid belt around a stellar-mass object, we focus on the prediction of time--frequency drifting and polarization. In this scenario, the frequency drifting is mainly caused by the geometric structure of a pulsar magnetosphere, and the drifting rate--frequency index is found to be $25/17$. On the other hand, by considering the typical differential mass distribution of incident asteroids, we find that an asteroid with mass $m\gtrsim 10^{17}~{\rm g}$ colliding with the pulsar would contribute abundant gravitational energy, which powers an FRB. A broad frequency band of the FRBs would be expected, due to the mass difference of the incident asteroids. In addition, we simulate the linear polarization distribution for the repeating FRBs, and constrain the linear polarization with $\gtrsim$ 30$\%$ for the FRBs with flux of an order of magnitude lower than the maximum flux.
	\end{abstract}
	
	
	\section{introduction}
	Fast radio bursts (FRBs) are extragalactic radio transients with millisecond durations and large observed dispersion measures (DMs) \citep{Lorimer2007,Keane2012,Thornton2013,Champion2016,Shannon2018,Platts2019,Petroff2019}, but their physical origin remains a mystery.
	So far, there are more than 100 FRBs that have been observed, and a small fraction of them show repeating behaviors \citep{spitler2016,CHIME2019a,CHIME2019b,Kumar2019,Fonseca2020}.
	The first reported repeating source, FRB 121102, was found to be localized in a dwarf galaxy with an extremely large observed rotation measure \citep{Chatterjee2017,Michilli2018}, and exhibits an intriguing downward structure that was then called the sub-pulse time--frequency drifting pattern \citep{Hessels2019,Josephy2019,Caleb2020}.

	The sub-pulse drifting patterns were found in most of the repeating sources rather than any non-repeating FRBs, suggesting that they are very likely to be a common property for repeaters. A lot of efforts have been made to explain the drifting pattern. Propagation effects seem to be possible, for instance, plasma lensing \citep{Cordes2017}, which is expected to show both upward and downward drifts. On the other hand, if the coherent radiation of an FRB is generated in the pulsar magnetosphere\footnote{The other kind of models in which an FRB is generated by energy dissipation via an outflow interacting with an ambient medium, i.e., ``far-away'' models, have been proposed in the literature \citep[e.g.,][]{Lyubarsky2014,Beloborodov2017,Beloborodov2019,Ghisellini2017,Waxman2017,Gruzinov Waxman2019,Metzger2019,Margalit2020b}.} \citep{Cordes2016,Kumar2017,Ghisellini 2018,Katz2018,Yang2018,Lu Kumar2019,Wang2019,Wang2020,Lu2020,Yang2020}. Moreover, the recent work by \cite{Beniamini2020} shows that the light-curve variability timescales and the spectro-temporal correlations in the data would provide strong evidence in favor of the  ``close-in'' models in which an FRB is proposed to occur near a magnetar.
	\citet{Wang2020} found that there is a small probability to generate the upward drifting but in most cases the downward drifting appears. Therefore, we mainly account for the downward drifting in this work.
	
	Polarized observation is useful for identifying the origin of FRBs.
	The linear polarization component is almost 100$\%$ for FRB 121102 \citep{Michilli2018,Gajjar2018} and is also quite high for other repeating sources \citep{CHIME2019b,Fonseca2020,Day2020}.
	For curvature radiation, electromagnetic waves are polarized in the plane of the magnetic field lines. The flux reaches its maximum when the angle between the electron trajectory plane and line of sight (LOS) is limited within 1/$\gamma$, and the polarization degree is almost 100$\%$. Besides, strong linear polarization can be generated by a magnetized plasma in the magnetospheres of pulsars \citep[e.g.][]{Melrose1979}.
	
	The periodic activity could provide a new channel for studying the nature of FRBs.
	A periodicity $\sim$16 day \citep{CHIME2020a} and a possible periodicity $\sim$159 day \citep{Rajwade2020} were recently detected to arise from FRB 180916.J0158+65 and FRB 121102, respectively.
	The periodicity was proposed to possibly result from
	precession of a magnetar \citep{Levin2020,Zanazzi2020}, orbital precession of a binary \citep{YangZou2020}, spin of an ultra-long-period magnetar \citep{BeniaminiWadiasingh2020} or 
	orbital movement of a pulsar and a stellar-mass companion in which case some properties of an extragalactic asteroid belt can be well constrained \citep{dai 2020}. Subsequently, this model was further argued by invoking a magnetar-asteroid impact \citep{dai 2020b} to account for FRB 200428 \citep{CHIME2020b,Bochenek2020} associated with an X-ray burst from the Galactic magnetar SGR 1935+2154 \citep{Li2020,Tavani2020,Mereghetti2020,Ridnaia2020}. A several-millisecond time delay of the two peaks of the X-ray burst with respect to the two pulses of FRB 200428 was found \citep{Li2020,Mereghetti2020}. This time delay is on the contrary of the trend naively expected in both blast-wave models and magnetospheric models in which an FRB is triggered by Alfv\'enic waves from the magnetar's surface \citep[for a discussion see][]{Margalit2020a}. Thus, this observation seems to favor the model of \cite{dai 2020b}.

	In this paper, we propose that the sub-pulse drifting pattern is produced by the motion of emitting bunches from an asteroid along the magnetic field lines around the Alfv\'en radius, and predict that the linear polarization distribution is peaked near 100$\%$. This paper is organized as follows. In Section \ref{sec2}, we demonstrate the frequency drifting generation in our model. In Section \ref{sec3}, we explain the polarization of repeating FRBs. A discussion is shown in Section \ref{sec4} and our summary is given in Section \ref{sec5}.
	The convention $Q_x$ = $Q/10^x$ in cgs units is used throughout this paper.
	
	\section{Time--Frequency Drifting}\label{sec2}
	
	In order to let an asteroid fall freely over the stellar surface, a pulsar with a slow rotation and an old age is required. In this scenario, a low cooling luminosity and an extremely low spin-down power would cause evaporation and ionization to become ineffective for the asteroid falling into the magnetosphere of the pulsar \citep{Cordes&Shannon2008}.
	\begin{figure}[!htbp]
		\centering
		\includegraphics[width=0.48\textwidth]{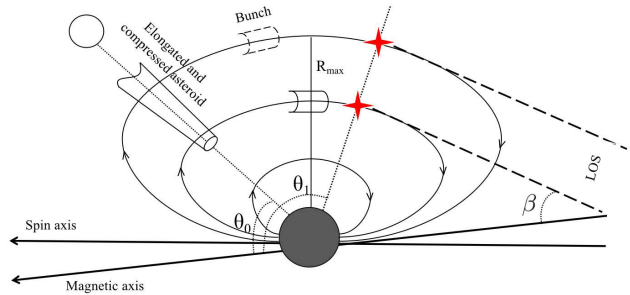}
		\caption{Schematic geometry of collision between an elongated, compressed asteroid and a pulsar, producing sub-burst downward frequency drifting of repeating FRBs. As the asteroid gradually falls with the immediate emergence of numerous net charges, a stray electric field outside of the asteroid has a component parallel to the magnetic field, which causes abundant bunches to leave from the surface of the asteroid and move along the magnetic field lines. A higher-frequency signal is generated closer to the pulsar due to a smaller curvature radius. The dashed lines display the line of sight (LOS) and $\theta _0$ is the angle between the radial direction of the asteroid infalling and the magnetic axis. When the bunches move along the magnetic field lines sweeping across the LOS, the angle between the spark (marked by red stars) and the magnetic axis is represented by $\theta_1$. $\beta$ denotes the angle between the LOS and the magnetic axis.}
		\label{Fig1}
	\end{figure}
	
	Figure \ref{Fig1} shows a higher frequency radio signal from a bunch observed earlier in a lower height, and vice versa. Coherent curvature emission from lots of bunches of different magnetic field lines can be observed. The curvature radius $\rho_{c}$ and the distance along the magnetic field lines ($\Delta l$) of the sub-pulses of high frequency and low frequency are different. The emission timescale of bunches is of order nanosecond because of the frequency of $\sim$ GHz which is extremely shorter than millisecond-duration radio bursts, so a large number of bunches are required to sweep across the LOS \citep{Yang2018}.

	\subsection{Heating and Evaporation Mechanism of Asteroids \label{sec2.1}}
	
	During each pulsar - asteroid collision, the heating and evaporation mechanism of an asteroid can determine whether the asteroid can be injected into the magnetosphere of an old-aged pulsar. Evaporation occurs inside the light cylinder and the evaporation temperature of an asteroid changes from 1500 to 2600 K \citep{cheng1985}. On the other hand, the asteroid temperature is determined by an equilibrium between the cooling rate and the heating rate. For an asteroid in thermal balance, due to surface radiation of the neutron star, the asteroid temperature $T_a$ is given by $T_a\simeq T_{\rm NS}\left(R_{\rm NS}/2 R\right)^{1/2}\simeq  707T_{\rm NS,5}R_{10}^{-1/2} \;{\rm K}$ \citep{Cordes&Shannon2008}, where $T_{\rm NS}$ and $R_{\rm NS}$ are the surface temperature and the radius of the neutron star, respectively. The cooling mechanism is dominated by the stellar surface blackbody radiation, and the surface temperature of the old-aged neutron star reaches $\sim$ 10$^5$ K \citep{Shapiro 1983}. Their evaporation rates are too low to provide the full current density for $R\sim 10^{10}{\rm cm}$. Therefore, the asteroid cannot be evaporated outside of the light cylinder. As the asteroid reaches the surface of the neutron star, numerous net charges near the surface of the asteroid can be contributed instantly by the separation of electrons and ions in the interior of the asteroid due to reverse Lorentz forces. It's dominated by runaway ohmic heating because an induced electric field can provide sufficient current through the asteroid. As the asteroid crosses the magnetic field lines transversely, an induced electric field, $\boldsymbol{E_1} = -(\boldsymbol{\Omega}\times \boldsymbol{R})\times \boldsymbol{B}/c$, can be produced, where $\Omega$ is the pulsar's angular velocity. When the condition of the induced azimuthal magnetic field at the asteroidal surface $2\pi r J/c \lesssim B$ can be satisfied, where $J$ is the current density, the ohmic heating power is given by $\widetilde{P}\approx r^3 J E$ \citep{Cordes&Shannon2008}. Thus, the maximum current density can be estimated $J_{\rm max}\approx c B/2\pi r$. The maximum heating power can be estimated as
	\begin{eqnarray}
	\widetilde{P}_{\rm max}
	&\simeq& \left (\frac{r B_{\rm p} R_{\rm NS}^3}{ P^{1/2}}\right )^2\left (\frac{1}{2\pi R}\right )^5 \nonumber\\
	&\simeq& 10^{33}  \;{\rm erg}\;{\rm s}^{-1}\; r^2_{4}B^2_{{\rm p},14}R^6_{\rm NS,6}	P^{-1}_0R^{-5}_7,
	\end{eqnarray}
	and thus the asteroidal temperature by runaway ohmic heating is
	\begin{eqnarray}
	T_a &\simeq& \sigma \left (\frac{\widetilde{P}_{\rm max}}{4 \pi \sigma_{\rm SB} r^2}\right )^{1/4}\nonumber \\
	&\simeq& 2600 \sigma_{-4} B^{1/2}_{{\rm p},14}R^{3/2}_{\rm NS,6}P^{-1/4}_{0}R^{-5/4}_7\; {\rm K},
	\end{eqnarray}
	where $\sigma_{\rm SB}$ is the Stefan-Boltzmann constant, $c$ is the speed of light, $r\sim  10^4$ cm denotes the radius of the stretched asteroid, and $B_{\rm p} = 10^{14}$ G is the surface magnetic field of the pulsar.
	If the thermal conversion efficiency $\sigma \approx 2\times10^{-4}$ , then the pulsar's rotation period can be limited to be $P \gtrsim 1 $s when the asteroidal surface reaches a maximum temperature 2600 K with the location of the asteroid $\sim 10^7$ cm.
	
	\subsection{The Downward Drift Pattern \label{sec2.2}}
	\begin{figure}
		\centering
		\includegraphics[width=0.48\textwidth]{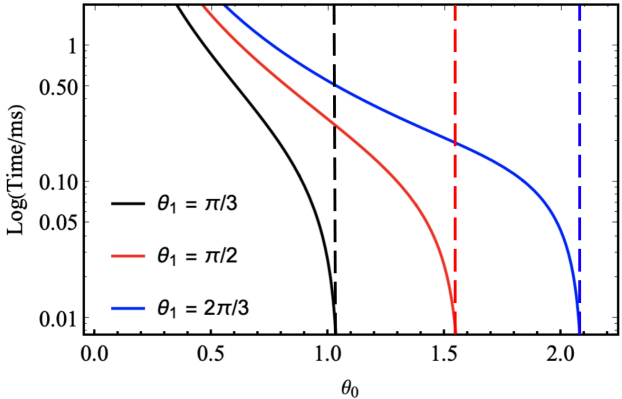}
		\caption{The time delay as a function of $\theta_0$ from Equation (\ref{eq10}). The lines in black, red and blue indicate $\theta_1=\pi/3$, $\pi/2$ and $2\pi/3 $, respectively and the corresponding dashed lines denote $\theta_0=\pi/3,\pi/2$ and $2\pi/3$, respectively. The initial distance between the high and low frequency bunches are the length of the asteroid $\sim 10^7$cm. }
		\label{Fig2}
	\end{figure}

	The free-falling asteroid passing through the magnetosphere of the pulsar can generate an induced electric field ${\bf E_2}$, which is shown as a force-free field inside the asteroid \citep{dai 2016}.
	This electric field outside of the asteroid has a component parallel to the magnetic field, making electrons dragged out of the surface of the asteroid and then accelerating these electrons to relativistic energies. These dragged electrons form bunches of charged particles, which give rise to coherent curvature radiation in the inner magnetosphere of the neutron star.
	
	Generally, the phenomenon of time--frequency drifting can be interpreted by the geometrical effect within the framework of curvature radiation \citep{Wang2019,Wang2020}, since the typical frequency $\nu$ strongly depends on the curvature radius $\rho_c$, i.e., $\nu=3c\gamma^3/(4\pi\rho_c)$, where $\gamma$ is the Lorentz factor.
	As the radially elongated, transversely compressed asteroid falls down to the radius $R$ from the pulsar and the angle $\theta_0$ between the asteroidal falling direction with the magnetic axis.
	According to the magnetic dipole geometry, bunches leave from the surface of the asteroid and then they pass through the distance along the magnetic field lines $\Delta l = v_e \Delta t$, where $\Delta l$ is given by the Appendix (bunches' motion nearly at the speed of light).
	When the bunches move to ($R_1$, $\theta_1$), their emission can be observed by us (see Figure \ref{Fig1}).
	In this case, the bunches at a more-curved part of the magnetic field lines are always seen earlier, thus emitting higher-frequency waves.
	The difference of the distance along the different magnetic lines is given in the Appendix, and thus the delay of the high frequency bunches relative to the low frequency bunches is given by
	\begin{equation}
	dt =\frac{d R}{c \sin^2 \theta_0 }\left[Z(\theta_0,\theta_1)-\cos(\beta-\theta_1)\sin^2 \theta_0\right],\label{eq10}
	\end{equation}
	where
	\begin{equation}
	Z(\theta_0,\theta_1)=\int_{\theta_0}^{\theta_1} \sqrt{1+3\cos ^2\theta}\sin \theta\, d\theta,
	\end{equation}
	where $\beta$ denotes the angle between the LOS and the magnetic axis.
	As shown in Equation (\ref{eq10}), the time delay mainly depends on the separation in the radial direction.  The relation between  $\theta_1$ and  $\beta$ can be written as \citep{Lyutikov2020}
	\begin{equation}
	\cos2\theta_1=\frac{1}{6} \left[\sqrt{2}\cos\beta\sqrt{\cos(2\beta)+17}+\cos(2\beta)-1\right].
	\end{equation}
	
	The time delay as a function of $\theta_0$ is shown in Figure \ref{Fig2}. As shown in Figure \ref{Fig2}, we set $\theta_1$ = $\pi/3$, $\pi/2$, and $2\pi/3$, respectively, the difference between the dashed line and the solid line corresponds to $\theta_1-\theta_0$ for the observed time delay.  The asteroidal length can be given by $l=l_{0}\left(R / R_{\rm b}\right)^{-1 / 2}$ for $R \leq R_{\rm i}$ \citep{Colgate1981}, where $R_{\rm b}$ and $R_{\rm i}$ denote the asteroids distorted tidally by the magnetar at breakup radius and the radius at which compression begins, respectively. The length of the asteroid can be estimated as $l\simeq  9.6 \times 10^{6}\mathrm{cm}\;m_{18}^{4 / 9} \rho_{0,0.9}^{-5 / 18} s_{10}^{-1 / 6}\left(M / 1.4 M_{\odot}\right)^{1 / 6} R_{7}^{-1 / 2} $, where $\rho_0$ and $s$ are the iron-nickel asteroid's original mass density and tensile strength, respectively. If the distance between the bunches of high and low frequency from the surface of the asteroid is too close, no matter what angle of the asteroid falls, there is no delay effect in the direction of the LOS. There is always a low limit of the incident angle for the time delay of millisecond order between high and low frequencies to be observed in a given LOS. However, if $\theta_0$ is larger than $\theta_1$ (see Figure \ref{Fig1}), then we are unable to observe the frequency drift.
	
	For a magnetic dipole configuration, from a geometric point of view, during the asteroid's fall, the bunches leave the asteroidal surface and move along the magnetic field line to sweep across the LOS, which can naturally produce sub-pulses with downward time--frequency drifting. Assuming that the Lorentz factor is a constant, we have
	\begin{equation}
	\frac{d\nu}{dt}=-\frac{  4\pi C(\theta)  }{9\gamma^3 }\nu^2,\label{eq00}
	\end{equation}
	where $C(\theta)$ is given by
	\begin{equation}
	C(\theta)=\frac{  \sin\theta_1  \sin^2\theta_0 (1+3\cos^2\theta_1)^{3/2}}{(1+\cos^2\theta_1) \left[Z(\theta_0,\theta_1)-\cos(\beta-\theta_1) \sin^2 \theta_0\right]}.
	\end{equation}
	This result deviates from the observation of FRB 121102 \citep{Josephy2019}, which shows a linear tendency. In the following discussion, we consider that the Lorentz factor evolves with $R$.
	
	The coherent radiation leads to rapid cooling of emitting particles and therefore requires continuous acceleration of the bunched particles.
	Generally, the cooling timescale of curvature emission in the observer's rest frame is given by $t_{\rm cool}=27m_e c^3 \gamma^3/16\pi^2 e^2 \nu^2 N_e$, where $N_e$ is the number of electrons in the bunch \citep{Kumar2017}. Then the parallel electric field can be written as
	\begin{equation}
	E_\|=\frac{16\pi^2eN_e[9\sin\theta_1(1+\cos^2\theta_1)]^{2/3}}{27c^{4/3} [4\pi (1+3\cos^2\theta_1)^{3/2}]^{2/3}}\frac{\nu^{4/3}}{R^{2/3}}.\label{cooling}
	\end{equation}
	Owing to the fact that the induced electric field $\boldsymbol{E_{ 1}}$ is perpendicular to the magnetic field line, so it cannot accelerate electrons.
	While the asteroid travels through the magnetic field lines longitudinally, causing the induced electric field $\boldsymbol{E_{ 2}}$ to accelerate electrons \citep{dai 2016}, we have
	\begin{eqnarray}
	\boldsymbol{E_{ 2}}=-\boldsymbol{v_{\rm ff}}\times\boldsymbol{B}/c &=& 2\times 10^{13} {\rm volt\;cm^{-1}}\boldsymbol{e_2}\nonumber\\
	&\times &\left(\frac{M}{1.4M_\odot}\right)^{1/2}\mu_{32}R^{-7/2}_7 ,
	\label{eqind}
	\end{eqnarray}
	where $\mu=B_{\rm p} R_{\rm NS}^3$ denotes the magnetic dipole moment of the pulsar, $v_{\rm ff}$ is the asteroidal free-fall velocity, which can be written as  $v_{\rm ff}=(2GM/R)^{1/2}$, and $\boldsymbol{e_2}$ is the unit vector of $\boldsymbol{E_{ 2}}$.
	The cooling electric field is equal to the induced electric field to prevent rapid cooling of electrons. From Equations (\ref{eq10}), (\ref{cooling}) and (\ref{eqind}), the drifting rate is thus given by
	\begin{equation}
	\frac{d\nu}{dt}=-1.9\left(\frac{eN_e}{\mu}\right)^{6/17}\frac{\pi^{8/17}c^{15/17}}{(GM)^{3/17}}D(\theta)\nu^{25/17},\label{eq000}
	\end{equation}
	where $D(\theta)$ is given by
	\begin{equation}
	D(\theta)=\frac{\sin^2 \theta_0 \left[\sin\theta_1(1+\cos^2\theta_1)  \right]^{4/17}}{ (1+3\cos^2\theta_1)^{6/17}\left[Z(\theta_0,\theta_1)-\cos(\beta-\theta_1)\sin^2 \theta_0 \right]  } .
	\end{equation}
	The frequency drifting rates are negative, i.e., downward drifting, matching the observations of most repeating FRBs \citep{Hessels2019,Josephy2019,CHIME2019a,CHIME2019b,Fonseca2020,Caleb2020}.
	
\begin{figure}[!htbp]
		\centering
		\includegraphics[width=0.48\textwidth]{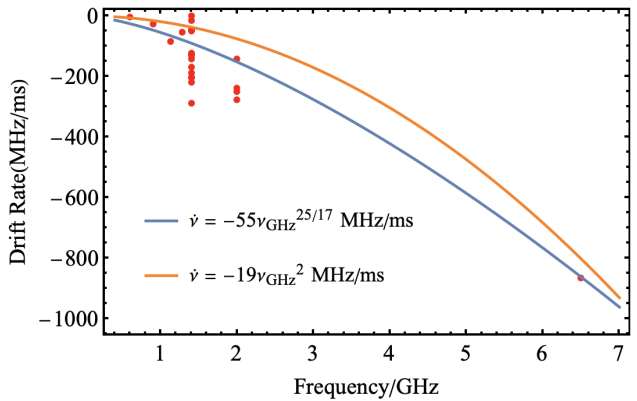}
		\caption{The drifting rate as a function of frequency. The red dots represent the observations of FRB121102 \citep{Hessels2019,Josephy2019,Caleb2020}. The blue line implied by Equation (\ref{eq000}) shows a fitting curve to the observed data in our model. The orange line inferred from Equation (\ref{eq00}). We adopt the physical parameters: $\theta_0=\pi/6$, $\theta_1=\pi/3$, $\beta=\pi/2$.
		}
		\label{fig3}
	\end{figure}

	In Figure \ref{fig3}, we present the drifting rate as a function of the central frequency, and the fitted parameter $N_e\sim 10^{27}$ matches the electron number of \cite{dai 2016}, which indicates that our model is self-consistent.  As shown in Figure \ref{fig3} , the drift rate implied by  Equation (\ref{eq000}) is more consistent with the observations than Equation (\ref{eq00}). The drifting rate at a higher frequency is larger than that at a lower frequency. Sub-pulses at the same central frequency may have fluctuated the drifting rate. This fluctuation may be related to the fluctuation of the emitting electron Lorentz factor and the angle of the asteroidal falling due to a random-falling asteroid.

	\section{Polarization}\label{sec3}
	
	Polarization is an important clue to the radio radiation mechanism and related to radiation geometry.  At present, polarization of dozens of FRBs has been observed\footnote{The FRB Catalogue http://www.frbcat.org.\label{note1} }. There are completely unpolarized FRBs that have been observed (e.g. FRB 150418 \citep{Keane2016}). FRB 140514 displays only circular polarization \citep{Petroff2015}, a few show only linear polarization (e.g. FRB 121102 and FRB 180916.J0158+65 \citep{Michilli2018,CHIME2019b}), and several FRBs show both (e.g. FRB 110523 \citep{Masui2015}). FRB 121102, FRB 180916.J0158+65, FRB 190604 and FRB 190303 \citep{Fonseca2020} are repeating sources. The linear polarization is very high, nearly 100$\%$ linearly polarized for the first three FRBs.
	Especially, the linear polarization of FRB 190303 was constrained to be $\geq$ 20$\%$ \citep{Fonseca2020}, which could suggest the emission mechanism of the FRBs.

	For curvature radiation, the electron trajectory is almost along a magnetic field line. We define the radiation angular frequency as $\omega$, the angle between the LOS and the charge trajectory plane as $\chi$, and the Lorentz factor of relativistic electrons as $\gamma$. In the direction of the electron's motion, the radiation would be beamed in a narrow angle 1/$\gamma$. The energy per unit frequency interval per unit solid angle is given by \citep[e.g.][]{jackson 1975}
	\begin{equation}
	\begin{split}
	\displaystyle\frac{dI}{d\Omega d\omega} =\displaystyle\frac{e^2}{3\pi^2 c} \left ( \displaystyle\frac{\omega \rho_c }{c}\right ) ^2 \left ( \displaystyle\frac{1}{\gamma^2}+{\chi^2}\right ) ^2 \times
	\\
	\left [ K^2_{2/3}(\xi)+  \displaystyle\frac{\chi^2}{(1/\gamma^2)+\chi^2}K^2_{1/3}(\xi)\right ],
	\end{split}
	\end{equation}
	where $ K_{\nu}(\xi$) denotes the modified Bessel function, $\xi =(1/\gamma ^{2}+\chi ^{2})^{3/2} \omega \rho_c/3c$, the first term corresponds to the parallel component in the trajectory plane and the second term corresponds to the polarized component that is perpendicular to the LOS. The degree of linear polarization is given by \citep{jackson 1975}
	\begin{equation}
	\Pi= \left|
	\frac{\chi^{2} K_{1/3}^{2}(\xi)-\left(\frac{1}{\gamma^{2}}+\chi^{2}\right) K_{2/3}^{2}(\xi)}{\chi^{2} K_{1 / 3}^{2}(\xi)+\left(\frac{1}{\gamma^{2}}+\chi^{2}\right) K_{2/3}^{2}(\xi)}
	\right| \label{eq9}.
	\end{equation}    In the following calculation, we adopt the parameters: $\gamma=200$ and $\chi$ changes from 0 to $1/\gamma$. Therefore, we can obtain the relation between curvature radiation of electrons and the degree of the linear polarization.
	\begin{figure}[!htbp]
		\centering
		\includegraphics[width=0.48\textwidth]{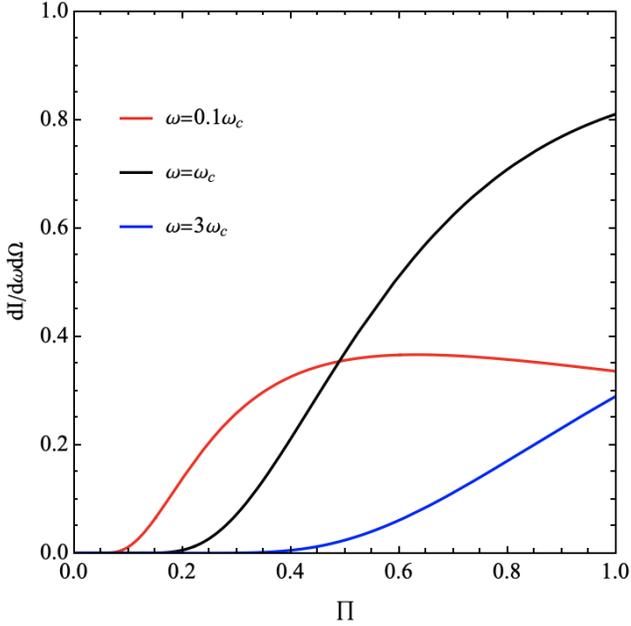}
		\caption{The curvature radiation of electrons as a function of the degree of the linear polarization. The unit of $dI/d\Omega d\omega$ is arbitrary and the parameter $\chi$ ranges from 0 to 10/$\gamma$. The red, black and blue lines denote the angular frequency  $\omega=0.1\omega_c$, $\omega_c$ and $3\omega_c$, respectively, where $\omega_c$ is the characteristic frequency.}\label{Fig4}
	\end{figure}
	In Figure \ref{Fig4}, we can see that the energy radiated per unit frequency interval per unit solid angle increases with the linear polarization for $\omega \geq \omega_c$. It's noted that $dI/d\Omega d\omega$ increases firstly and then decreases slowly with $\Pi$ at the high linear polarization for $\omega < \omega_c$, since the flux would reach the maximum when $\chi$ deviates the LOS slightly.
	If the observed flux varies within an order of magnitude for different sub-bursts, the linear polarization would be $\gtrsim$ 30$\%$ for the characteristic frequency, which is nearly consistent with the observed linear polarization constraint of the FRB 190303 \citep{Fonseca2020}. If $\omega\gg\omega_c$, the low limit of the linear polarization may be raised until nearly 100$\%$, while $\omega\ll\omega_c$, the linear polarization could be extended to the lower values.

\begin{figure}
		\centering
		\includegraphics[width=0.48\textwidth]{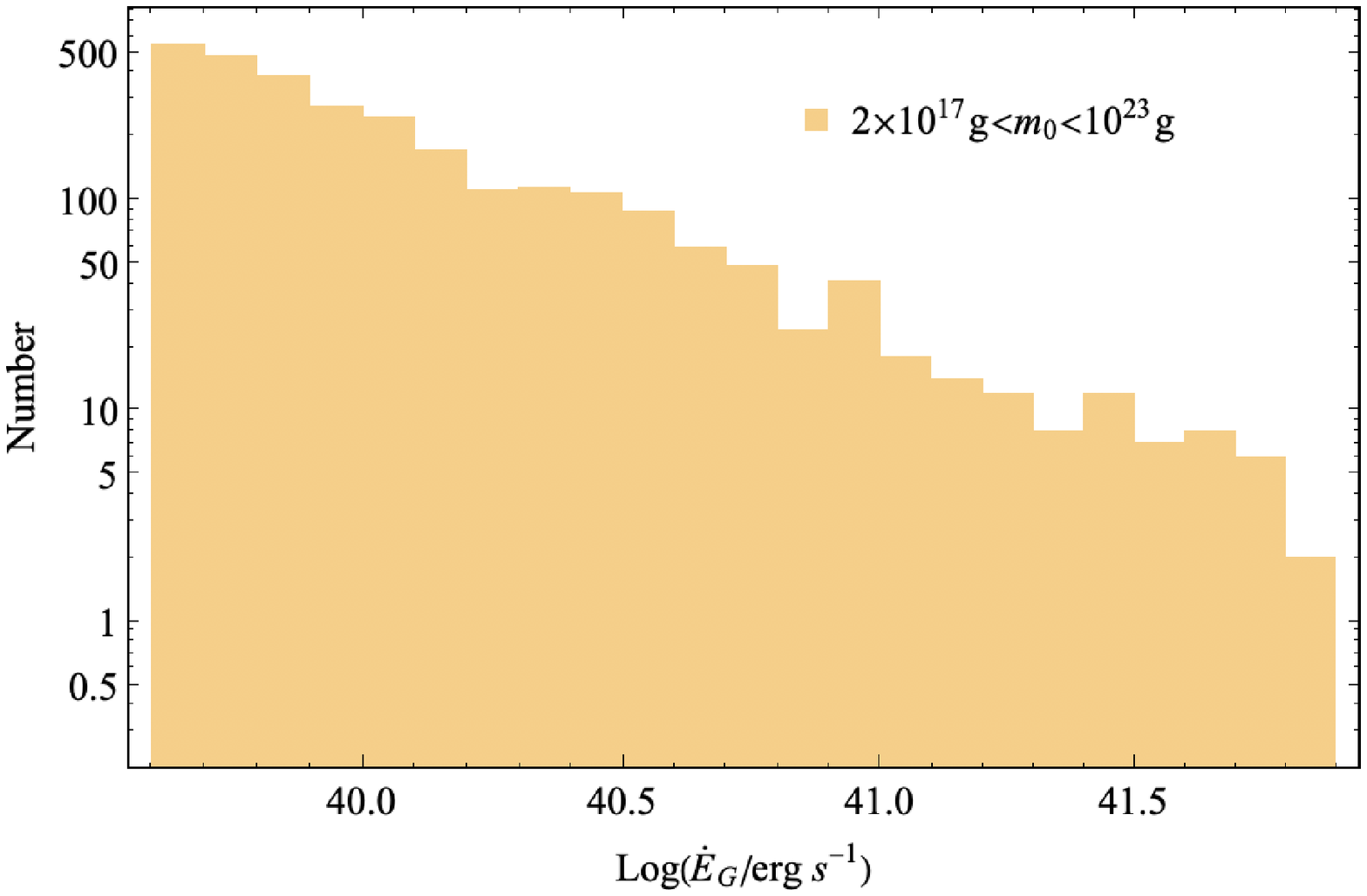}
		\includegraphics[width=0.48\textwidth]{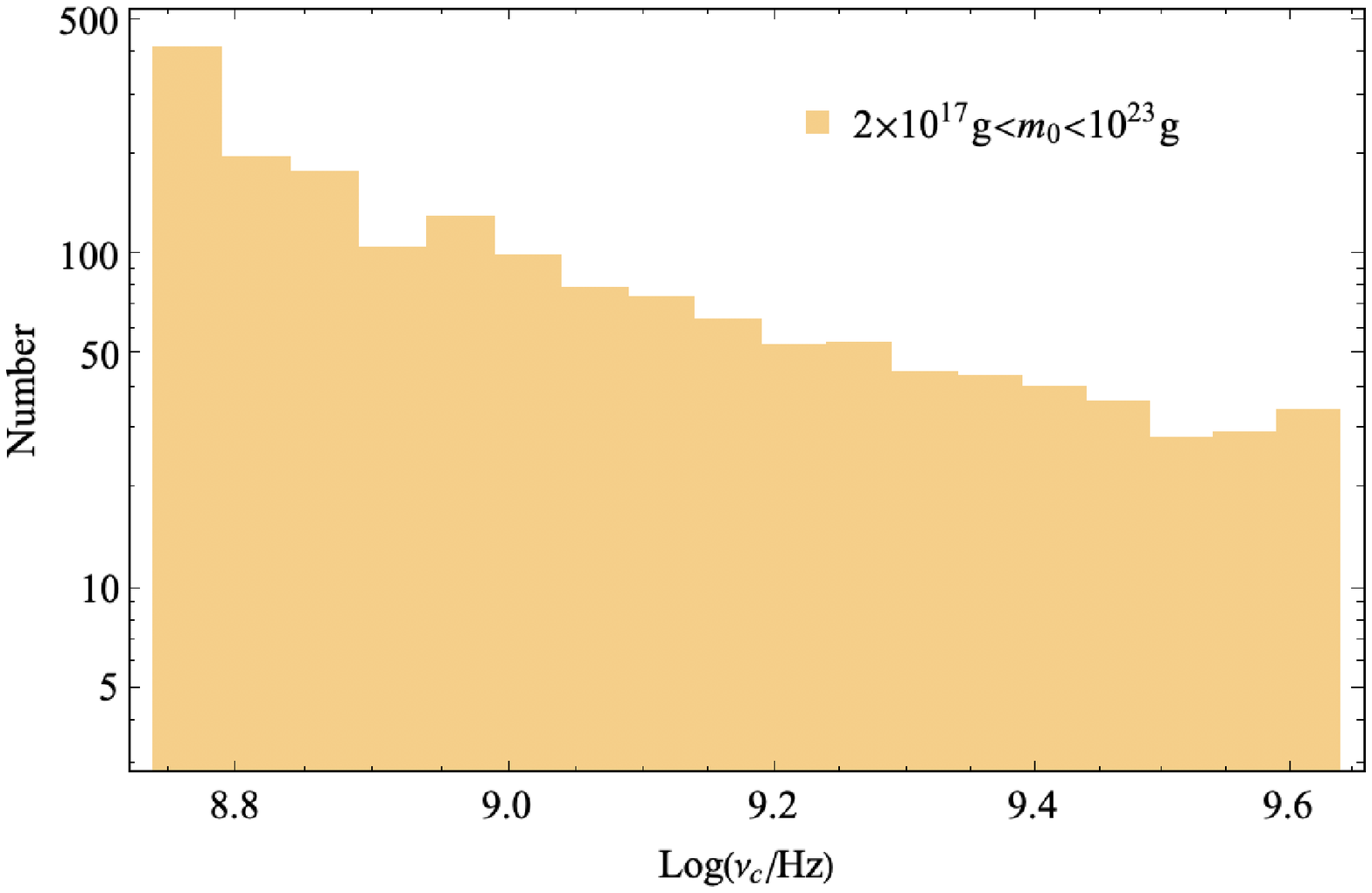}
		\includegraphics[width=0.48\textwidth]{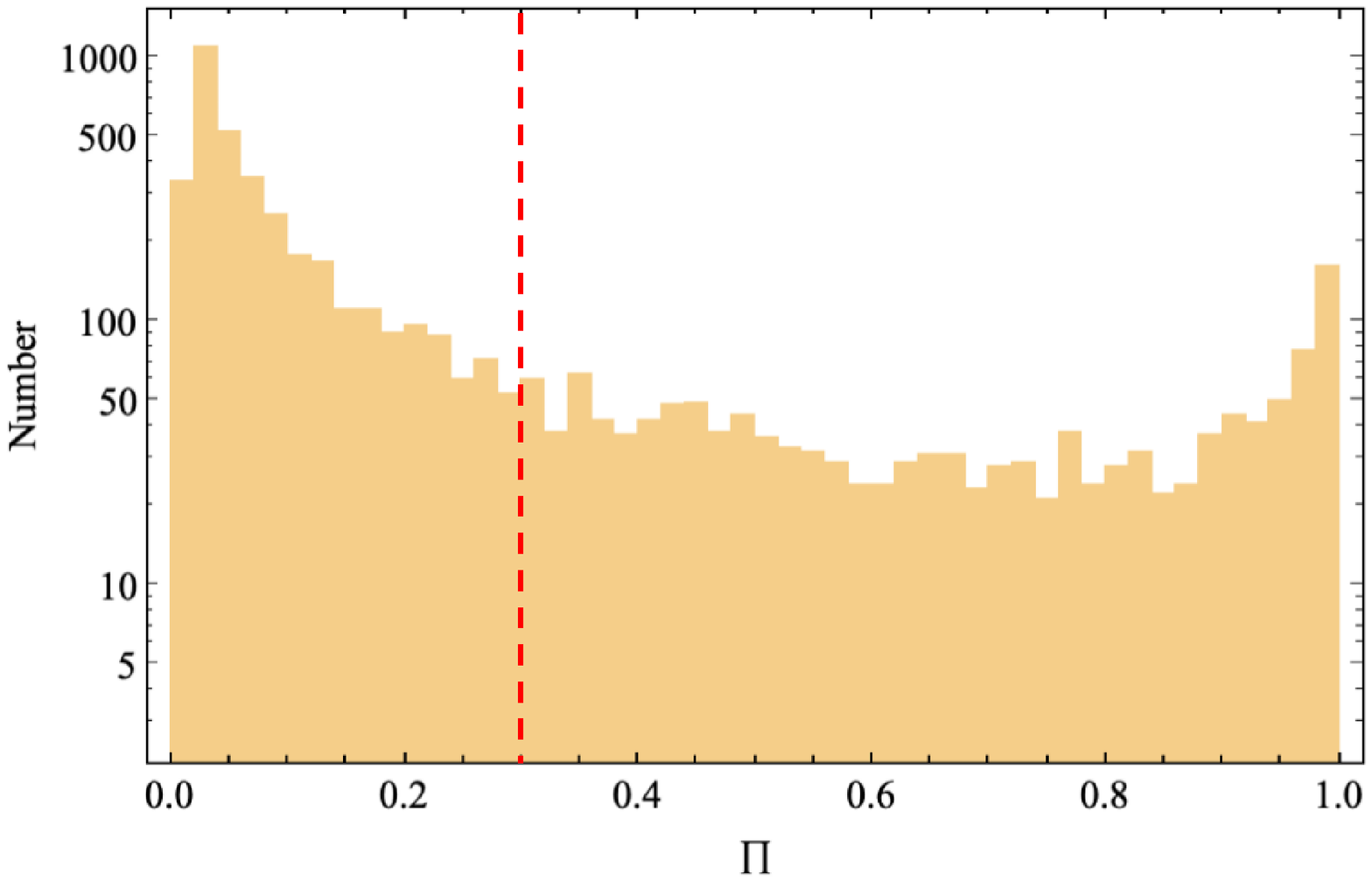}
		\caption{The upper, middle, and bottom panels denote the distributions of the gravitational energy, the characteristic frequency and the linear polarization, respectively. The range of the asteroidal mass for observable FRBs are $2\times 10^{17}{\rm g}<m<10^{23}$g and the corresponding  asteroid radius is  $3\times 10^5{\rm cm}<r_0<2.3\times10^7$cm. The red dashed line denotes the linear polarization of $\sim$ 30$\%$ is corresponding to the low limit of the flux within an order of magnitude variation.}\label{Fig5}
	\end{figure}

	As the asteroid falls, the kinetic energy density is equal to the magnetic energy density of the pulsar at the Alfv\'en radius, i.e.,
	\begin{equation}
	\frac{\rho v^2_{\rm ff}}{2}\sim\frac{\mu^2}{8\pi R^6},
	\end{equation}
	where $\rho\equiv m/\pi r^2 l$ is the mass density of the asteroid. The emission position could be related to the compressed asteroid radius. As the asteroid falls freely for $R<R_i$ (where $R_i$ is the radius where the asteroid begins to be compressed), the compressed asteroid length and radius evolves with $R$ can be given by \citep{Colgate1981}
	\begin{equation}
	l=l_0\left(\frac{R}{R_b}\right)^{-1/2},
	\end{equation}
	\begin{equation}
	r=r_0\left(\frac{R_i}{R_b}\right)^{1/4}\left(\frac{R}{R_i}\right)^{1/2},
	\end{equation}where $R_i=\kappa R_b=\kappa (\rho_0 r^2_0 GM/s)^{1/3}$, $R_b$ is the pulsar distorts tidally the asteroid at the breakup radius, $r_0$ is original cylindrical radius, $\kappa \equiv (5s/8\Lambda)^{2/5}\simeq 0.13 $, and $\Lambda$ is the solid body compressive strength \citep{Colgate1981}. Thus, the Alfv\'en radius can be given by
	\begin{eqnarray}
	R&=&2^{-11/27}(\pi s)^{1/27}\mu^{4/9}\kappa^{-1/9}(GM\rho_0)^{-7/27}r_0^{-2/27}\nonumber\\
	&=&1.38\times 10^7\;{\rm cm}(\kappa/0.13)^{-1/9} s^{1/27}_{10}\rho_{0,0.9}^{-7/27} \mu_{32}^{4/9}\nonumber\\
	& \times & r_{0,5}^{-2/27}(M/1.4 M_{\odot})^{-7/27}
	,
	\end{eqnarray}
	and the curvature radius at $R$ can be written as
	\begin{equation}
	\rho_c = \frac{R(1+3\cos^2\theta_1)^{3/2}}{3\sin\theta_1(1+\cos^2\theta_1)}.
	\end{equation}
	Following the differential size distribution of the solar system, the asteroid distribution of an extragalactic asteroid belt can be assumed to be given by $dN/dr_0 \propto r_0^{-\delta} $, $\delta=\delta_1=2.3$ for $r_0<r_{\rm br}$ and $\delta=\delta_2=4.0$ for $r_0>r_{\rm br}$, where $r_{\rm br}\simeq 3 {\rm km}$ \citep{Ivezic2001,Davis2002,Yoshida2007,Ryan2015}.  According to the asteroid radius distribution, assuming the initial density is constant (i.e., 8 g/cm$^{3}$), the asteroidal mass distribution can be given by $dN/dm \propto m^{-\epsilon} $, $\epsilon=2$ for $m>m_{\rm br}$, where $m_{\rm br}\simeq 2\times10^{17}{\rm g}$.
	Thus, we can get the distribution of the gravitational energy release rate via $\dot{E}_G\simeq GMm/R\Delta t$, where $\Delta t=1$ ms, $R\sim10^7$cm. While the observed minimum luminosity of the extragalactic FRBs $\sim$ $10^{39}$ erg/s (see footnote 2), the low limit of the incident asteroid mass and radius can be constrained to $m_{\rm min}\simeq5.4\times 10^{16}$g and $r_{0,\rm min}\simeq 1.8\times 10^{5}$cm, respectively. The observable FRBs could arise from $m>m_{\rm min}$ and $r_0>r_{0,\rm min}$. Since $m_{\rm br}>m_{\rm min}$, we can consider the asteroidal mass distribution for $m>m_{\rm br}$. From the top panel of the Figure \ref{Fig5} , which indicates that a heavier asteroid would cause a stronger gravitational energy release rate.

	We next consider the solid angle random distribution of the incident asteroid angle $\sin \theta$ (from 0 to 1) and the emission angle $\sin\chi$ (from 0 to $\sin\left(10/\gamma \right)$), and the asteroidal radius distribution for $r_0>r_{\rm br}$. We can get the distribution of the characteristic frequency. As shown in the middle panel of Figure \ref{Fig5}, it's suggested that the curvature radius affected by a different asteroidal radius would lead to a different characteristic frequency. A larger incident radius could lead to a lower Alfv\'en radius, causing a lower curvature radius and a higher observed frequency. Owing to the mass difference of the incident asteroids, FRBs with a broad frequency band would be observed. This is consistent with the detected frequency range from 400 MHz to 6.5 GHz for FRB 121102 \citep{Chatterjee2017,Michilli2018,Hessels2019,Josephy2019}.
	It can be seen from the simulation results (see the bottom panel of Figure \ref{Fig5} ) that a smaller emission angle $\chi$ would cause a higher linear polarization distribution due to the beaming effect.
	It is interesting that the polarization distribution is peaked at $\Pi\simeq0$ and $\Pi\simeq1$. The reason is as follows: for a single point source, the polarization decreases with $\chi$, which means that the polarization is maximum at $\chi=0$. However, by considering that the LOS is random, in most cases, one has $\chi\gg1/\gamma$, which makes the polarization become low (the corresponding flux is also much less than the maximum one). Therefore, the polarization distribution is peaked at $\Pi\simeq0$ and $\Pi\simeq1$ due to the two reasons mentioned above.

	\begin{figure}
		\centering
		\includegraphics[width=0.48\textwidth]{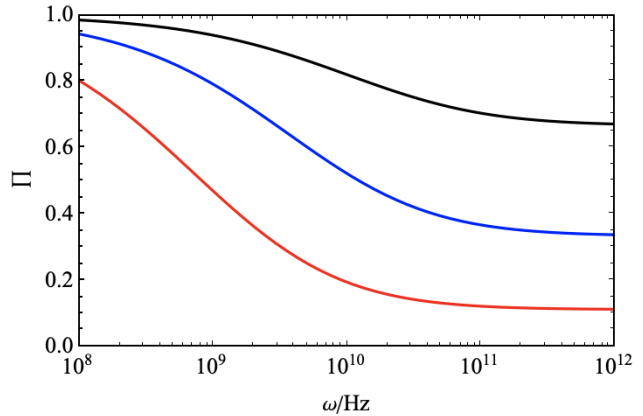}
		\caption{The linear polarization as a function of angular frequency. The black, blue and red solid line denote  $\chi$ = 1/2$\gamma$, 1/$\gamma$ and 2/$\gamma$, respectively. We adopt the physical parameters: $\gamma=200$, $\rho_c=10^7 {\rm cm }$.  }\label{Fig6}
	\end{figure}
	Figure \ref{Fig6} shows the linear polarization as a function of angular frequency. It can be seen that the linear polarization decreases with the angular frequency for a fixed $\chi$. Owing to the mass difference of incident asteroids, FRBs have a broad frequency band so that we can get the corresponding broad linear polarization.

	\section{Discussion}\label{sec4}
	Owing to the geometrical effect, bunches producing high-frequency emission from small curvature radii first arrive at an observer, and then bunches producing low-frequency emission are followed. Meanwhile, the NS rotates a small phase so that the magnetic configuration changes slightly. One can then calculate the time delay either via radial direction or rotational direction.
	
	Moreover, some sub-bursts have extremely short time intervals due to a short distance between high and low bunches from the asteroidal surface, i.e., one may not distinguish between the sub-bursts when extracting signals. The drifting rate's linear fitting index $\sim 1 $ \citep{Josephy2019} has a lot of uncertainties due to the drifting rate only observed in the 400-800 MHz and 6.5 GHz, respectively, mainly dominating the linear trend. On the other hand, the linear fitting curve does not cross the original point of the coordinate axis. In addition, the drifting rate detected at the central frequency of 1.4 GHz deviates from a Gaussian distribution. Since the drifting rate itself fluctuates, an observation of a wider frequency range ($\nu \gtrsim$ 8 GHz or $\lesssim$ 400 MHz) would provide significant information for understanding the relationship between frequency drifting rate and frequency,  which depends on the geometric distribution of $E_{\parallel}$.
	
	Some of bursts such as FRB 200428 and FRB 180916.J0158+65 exhibit upward drifting pattern \citep{CHIME2020a,CHIME2020b, Bochenek2020}. A long-range $E_{\parallel}$ and the time difference of bunch generation would lead to a time--frequency upward drifting pattern. If there is the upward frequency drifting, from the Figure \ref{Fig1}, it's suggested that the bunches would move a longer distance than those typically observed $\sim$ ms-order FRBs along the magnetic field lines. The long-range of an accelerating electric field may be triggered by the sudden distortion of the magnetic field lines around the asteroid or other globally accelerating electric fields such as $E_{\parallel}$ in the slot gap \citep{Muslimov2003,Muslimov2004}. Furthermore, similarly to \cite{Wang2020}, by considering bunches produced at different times, the radiation from a more curved part is seen later, which also provides a small possibility for the upward frequency drifting.
	
	As shown in the bottom panel of Figure \ref{Fig5}, we simulate the degree of linear polarization distribution for $r_0>r_{0,\rm min}$ and $\chi$ ranges from 0 to $\sin(10/\gamma)$. We find that the curvature radius would have a slight difference due to the radius difference of an incident asteroid, which is not sensitive to the linear polarization distribution. From Figure \ref{Fig4}, if the angular frequency is larger or smaller than the characteristic frequency, then the linear polarization could increase or decrease for FRBs with flux an order of magnitude lower than the maximum flux.
	
	The polarization angle swings, due to different magnetic field lines of the pulsar sweeping across the LOS, depend on the rotational period of the pulsar and the angle between the rotation axis and magnetic axis. In our model, the constraints on the slowly spinning pulsar and the small angle between the spin axis and magnetic axis can explain small variations in the polarization position angles for those repeating sources.

In our model, the observed spectra during the downwards drift can be understood as coherent curvature radiation by bunches \citep{Yang2018,Yang2020}. The observed light-curve variability timescale may be related to the difference between $\theta_0 $ and $\theta_1$. If the difference between $\theta_0 $ and $\theta_1$ is very small, then we can find the light-curve variability timescale is smaller than the ms time-scale, which provides strong evidence in favor of the magnetoshperic model. 
	
	\section{Summary}\label{sec5}
	
	In this paper, we have shown that the time--frequency downward drifting patterns are caused by the geometrical effect, i.e., the motion of emitting bunches from asteroids along the magnetic field lines at different heights. In our model, the production of FRBs requires that the pulsar is slowly-spinning and old-aged (i.e., the low spin-down power and cooling luminosity) so that the asteroids can enter its magnetosphere smoothly. Based on the unipolar inductor accelerating electric field balanced with the cooling electric field to prevent the electrons cooled quickly, we derived the frequency drifting rate index is 25/17 and the total radiating electron number $N_e\sim 10^{27}$. The frequency drifting doer not occur in the case that the asteroids enter the magnetosphere of a pulsar along the magnetic axis (i.e., $\theta_0 = 0$ or $\pi$). We found that the asteroidal incident angle would has a low limit for each given $\theta_1$ and the distance of high and low bunches from the asteroidal surface should be approximately the length of the asteroid since the time delay of ms-order between high and low frequencies is observed along the LOS. Furthermore, the variation of the drifting rate is probably a result of both the fluctuation of the electron number and the asteroidal random falling angle.

	In addition, we considered the asteroidal mass distribution and found that a heavier asteroid colliding with the pulsar could release a more abundant gravitational energy. FRBs with a broad frequency band would be expected due to the mass difference of the incident asteroids. We also simulated the linear polarization distribution for the repeating FRBs and found that the linear polarization can be constrained with $\gtrsim$ 30$\%$  for the FRBs with flux an order of magnitude lower than the maximum flux.

	\acknowledgments
	We are grateful to Xuelei Chen, Xue-Feng Wu, Yong-Feng Huang, Fa-Yin Wang, Jie-Shuang Wang and Jin-Jun Geng for helpful discussions and constructive comment. This work was supported by the National
	Key Research and Development Program of China (grant
	No. 2017YFA0402600) and the National Natural Science
	Foundation of China (grant No. 11833003). W.-Y. W. thanks the acknowledges the support of the NSFC Grants 11633004, 11653003, the CAS grants QYZDJ-SSW-SLH017, and CAS XDB 23040100, and MoST Grant 2018YFE0120800, 2016YFE0100300.

	\appendix\label{appendix}
	\section{Frequency drifting rate in the magnetosphere}
	For a dipole magnetic field configuration, the polar coordinates ($R$, $\theta$, $\phi$) and the radial direction position can be described as
	\begin{equation}
	R= R_{\rm max}\sin^2{\theta}\label{eq2},
	\end{equation}
	where $R_{\rm max}$ is the maximum distance of magnetic field lines across the equator.
	According to Equation (\ref{eq2}), $R_{\rm max} =R/\sin^2\theta$. Assuming the location of lower bunches and higher bunches at $R_1$ and $R_2$ to leave the surface of the asteroid. The distance along the magnetic lines can be written as respectively \citep{Yang2018}
	\begin{equation}
	\Delta l_1 =\frac{R_1}{\sin^2 \theta_0 } \int_{\theta_0}^{\theta_1} \sqrt{1+3\cos ^2\theta}\sin \theta\, d\theta ,
	\end{equation}
	\begin{equation}
	\Delta l_2 =\frac{R_2}{\sin^2 \theta_0 } \int_{\theta_0}^{\theta_1} \sqrt{1+3\cos ^2\theta}\sin \theta\, d\theta ,
	\end{equation}
	and the difference of the distance can be written as
	\begin{equation}
	d l =\frac{d R}{\sin^2 \theta_0 } \int_{\theta_0}^{\theta_1} \sqrt{1+3\cos ^2\theta}\sin \theta\, d\theta.
	\end{equation}
	The difference of the distance, $dl$, mainly depends on the separation in the radial direction $dR$. Relativistic electrons along the magnetic field lines can cause coherent curvature radiation for the Lorentz factor $\gamma$ and the curvature radius $\rho_c$, whose characteristic frequency $\nu$ of curvature radiation is
	\begin{equation}
	\nu = \displaystyle\frac{3c\gamma^3}{4 \pi \rho_c},\label{eq12}
	\end{equation}
	where the curvature radius $\rho_c$ at ($R$, $\theta$) is given by
	\begin{equation}
	\rho_c = \frac{R(1+3\cos^2\theta)^{3/2}}{3\sin\theta(1+\cos^2\theta)} .\label{eq13}
	\end{equation}
	The subscript $x$ and $y$ indicates the location of lower bunches and higher bunches sweeping across the LOS, respectively. We have
	\begin{equation}
	R_x=\frac{R_1\sin^2\theta_1}{\sin^2\theta_0},
	\end{equation}
	the curvature radius and the frequency at lower bunches are given by
	\begin{equation}
	\rho_x=\frac{R_1\sin\theta_1 (1+3\cos^2\theta_1)^{3/2}}{3\sin^2\theta_0 (1+\cos^2\theta_1)},
	\end{equation}
	and
	\begin{equation}
	\nu_x=\frac{9c\gamma^3 \sin^2\theta_0(1+\cos^2\theta_1)}{4\pi R_1\sin\theta_1(1+3\cos^2\theta_1)^{3/2}}.
	\end{equation}
	Similarly, the frequency at the higher bunches can be written in the same form,
	\begin{equation}
	\nu_y=\frac{9c\gamma^3 \sin^2\theta_0(1+\cos^2\theta_1)}{4\pi R_2\sin\theta_1(1+3\cos^2\theta_1)^{3/2}}.
	\end{equation}
	The difference of the frequency can be given by
	\begin{equation}
	d\nu=-\frac{9c\gamma^3 \sin^2\theta_0(1+\cos^2\theta_1)}{4\pi \sin\theta_1(1+3\cos^2\theta_1)^{3/2}}\frac{dR}{R^2},
	\end{equation}
	where we take the approximation $R\sim R_1 \sim R_2$. Thus, the frequency drifting rate is given by
	\begin{equation}
	\frac{d\nu}{dt}=-\frac{  4\pi \sin\theta_1  \sin^2\theta_0 (1+3\cos^2\theta_1)^{3/2} }{9\gamma^3 (1+\cos^2\theta_1) Z(\theta_0,\theta_1)}\nu^2.
	\end{equation}

\end{document}